\documentclass[twocolumn,floatfix,showpacs,amsmath,amssymb,superscriptaddress,prl]{revtex4}
\usepackage{graphicx}
\usepackage{dcolumn}
\usepackage{bm}

\begin{document}
\title{The nature of Feshbach molecules in Bose-Einstein condensates}
\author{T.~K\"ohler}
\affiliation{Clarendon Laboratory, Department of Physics, 
University of Oxford, Oxford, OX1 3PU, United Kingdom}
\author{T.~Gasenzer}
\affiliation{Institut f\"ur Theoretische Physik, Philosophenweg 16, 
69120 Heidelberg, Germany}
\author{P.~S.~Julienne}
\affiliation{Atomic Physics Division, National Institute of Standards and 
Technology, 100 Bureau Drive Stop 8423, Gaithersburg, Maryland 20899-8423}
\author{K.~Burnett}
\affiliation{Clarendon Laboratory, Department of Physics, 
University of Oxford, Oxford, OX1 3PU, United Kingdom}
\date{\today}
\begin{abstract}
We discuss the long range nature of the molecules produced in recent 
experiments on molecular Bose-Einstein condensation. The properties of these 
molecules depend on the full two-body Hamiltonian and not just on the states 
of the system in the absence of interchannel couplings. The very long range
nature of the state is crucial to the efficiency of production in the 
experiments. 
Our many-body treatment of the gas accounts for the full binary physics and 
describes properly how these molecular condensates can be directly probed.
\end{abstract}
\pacs{03.75.Kk, 34.50.-s, 36.90.+f, 05.30.-d}
\maketitle
Bose-Einstein condensation of molecules is an exciting challenge and 
opportunity in the physics of ultracold gases. As direct laser cooling of 
molecules is largely prevented by their densely lying rovibrational energy 
levels, several approaches now focus on the association of atoms 
in Bose-Einstein condensates. Present techniques 
are based on photoassociation 
\cite{Wynar00} 
or magnetic field tunable interactions 
\cite{Stenger99,Cornish00}.
The detection of molecular condensates, however, remains a demanding problem.
Recent experiments at JILA 
\cite{Donley02}
found evidence for molecular condensation by probing coherence properties of
the assembly of atoms plus molecules. 
The $^{85}$Rb condensate was exposed to a sequence of two fast magnetic field 
pulses in the vicinity of a Feshbach resonance. The pulses were separated by 
a variable period of time with a stationary magnetic field, termed the 
evolution period. The observed final densities of gas atoms, i.e.~the remnant
condensate and a ``burst'' of comparatively hot atoms, indicated a coherent 
coupling between atoms and diatomic molecules in a highly excited 
vibrational state 
during the evolution period. Several subsequent theoretical studies 
\cite{Kokkelmans02,Mackie02,KGB02} 
have concluded that the gas contained a molecular condensate at the end of 
the pulse sequence. The magnitude of the molecular fraction, and the 
precise mechanism of producing the molecules are a matter of continuing 
controversy 
\cite{molfield,Braaten}. 
To settle this issue the next generation of experiments could determine the 
coherent superposition of atomic and molecular components by directly 
detecting the molecules. The efficiency of different detection schemes based 
on laser excitation 
\cite{private}
depends sensitively on the number of molecules and the wave functions of the
diatomic bound states produced by a magnetic field pulse.

In this letter we shall give a full description of these wave functions for 
typical experimental magnetic field strengths and determine the molecular 
component in the evolution period of the pulse sequence. In particular, we 
shall show explicitly that the wave functions of the molecules have a spatial 
extent of the order of the scattering length. At the magnetic field strengths 
closest to resonance this length scale even becomes comparable with the mean 
distance between the atoms in the dilute condensate. Under these conditions a 
separation of the gas into atoms and diatomic molecules is physically 
meaningless. The strong binary correlations provided by the long range 
intermediate molecular states, however, are still crucial for the efficiency 
of the association of atoms to molecules. In the evolution period the typical 
static experimental field strengths are sufficiently far from the resonant 
field that the highly excited diatomic molecules can exist as a metastable 
entity of the gas. The microscopic many-body approach we apply to determine 
the molecular fraction 
\cite{KGB02,Koehler02}
accounts for the long range nature of the intermediate molecular states.
This approach treats bound and free molecular states formed during the pulse 
sequence in a unified manner and provides a straightforward understanding of 
the association of condensate atoms to molecules.

The experimental technique of magnetic field tunable interactions takes 
advantage of the Zeeman effect in the electronic energy levels of the atoms. 
In the JILA experiments 
\cite{Donley02}
the $^{85}$Rb condensate atoms were prepared in the $(F=2,m_F=-2)$ hyperfine 
state. Throughout this letter the open $s$-wave binary scattering channel of 
two asymptotically free atoms in the $(F=2,m_F=-2)$ state will be denoted as 
the $\{-2,-2\}$ open channel with an associated reference potential 
$V_\mathrm{bg}(r)$. When the $m_F$ degeneracy of the hyperfine levels is 
removed by an external magnetic field $B$ the potentials associated with the 
different asymptotic scattering channels are shifted with respect to each 
other. Although the $\{-2,-2\}$ open channel is only 
very weakly coupled to other 
open channels, it can be strongly coupled to closed channels.  
A zero-energy scattering resonance occurs when the field-dependent energy 
$E_\mathrm{res}(B)$ of a closed channel vibrational state 
(a Feshbach resonance level) $\phi_\mathrm{res}$ is 
tuned close to the dissociation threshold energy of $V_\mathrm{bg}(r)$. 
We note that the closed channel state $\phi_\mathrm{res}$ is not a
stationary state of the full two-body Hamiltonian, which has a coupling 
between the channels. If $E_\mathrm{res}(B)$ approaches the threshold from 
below, however, the overall potential matrix supports a shallow multi-channel 
bound state $\phi_\mathrm{b}$ with energy $E_\mathrm{b}$.
This proper stationary molecular state ceases 
to exist at the position of the resonance ($B_0=15.49\ $mT \cite{Donley02}),
for which the $s$ wave scattering length $a$ of two asymptotically free atoms 
in the $\{-2,-2\}$ open channel has a singularity. 
When $E_\mathrm{b}$ is sufficiently close to threshold, the component of 
$\phi_\mathrm{b}$ in the 
$\{-2,-2\}$ open channel then exhibits the universal form 
\cite{asymp}
$\exp(-r/a)/r$ at large 
relative distances of the two atomic constituents of the molecular state.
  
\begin{figure}[tb]
\begin{center}
\includegraphics[width=0.45\textwidth]{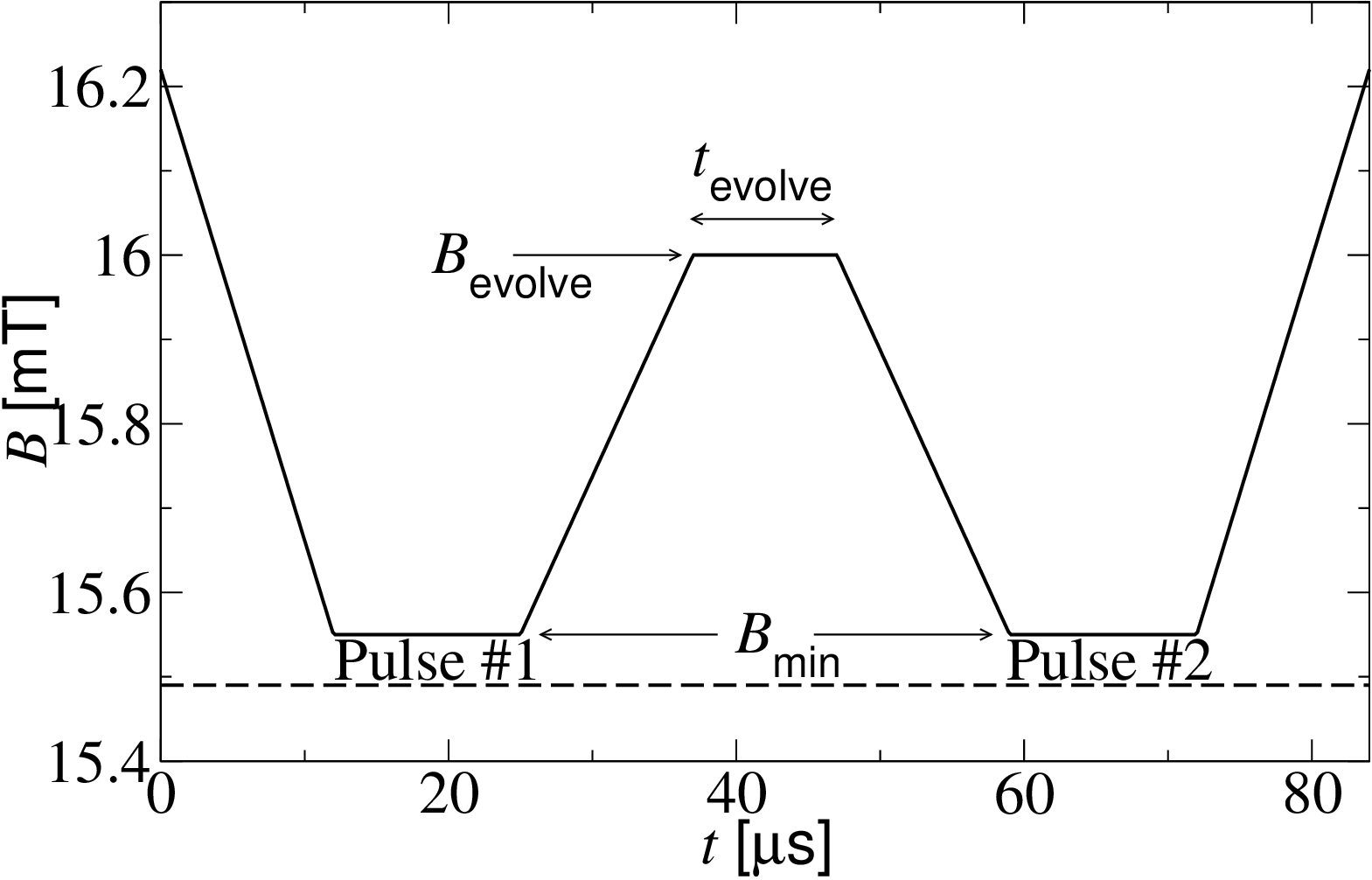}
\end{center}
\caption{Scheme of a typical magnetic field pulse shape in the low
density ($n_0=3.9\times 10^{12}\ \mathrm{cm}^{-3}$) experiments in 
Ref.~\cite{Donley02}. The minimum magnetic field 
strength of the first and second pulse is 
$B_\mathrm{min}=15.55\ $mT. In the evolution 
period the field strength is chosen as $B_\mathrm{evolve}=16.0\ $mT.  
In the course of the experiments the evolution time $t_\mathrm{evolve}$
as well as $B_\mathrm{evolve}$ were varied. The dashed line 
indicates the position of the resonance at $B_0=15.49\ $mT.}
\label{fig:pulse}
\end{figure}
We shall now describe more fully the long range nature of the molecular states 
for typical experimental magnetic field strengths in the fast sequence of 
pulses shown in Fig.~\ref{fig:pulse}. To this end we have reduced the complete 
multi-channel Hamiltonian of the relative motion of two atoms to a two-channel 
model:
\begin{align}
H_\mathrm{2B}=
\left(
\begin{array}{cc}
  -\frac{\hbar^2}{m}\nabla^2+V_\mathrm{bg}(r) & W(r)\\
  W(r) & -\frac{\hbar^2}{m}\nabla^2+V_\mathrm{cl}(B,r)
\end{array}
\right).
\label{H2B}
\end{align}
Here $m$ is the atomic mass of $^{85}$Rb. For the potential in the 
$\{-2,-2\}$ open channel we use a Lennard-Jones form
$V_\mathrm{bg}(r)=4\varepsilon\left[(\sigma/r)^{12}-(\sigma/r)^{6}\right]$ 
with $\sigma=37.3292\ a_\mathrm{Bohr}$ 
($a_\mathrm{Bohr}=0.052918\ $nm) and 
$4\varepsilon\sigma^{6}=C_6=4660\ $a.u.~\cite{Roberts01} 
(1 a.u. = 0.095734 yJ nm$^6$).
$V_\mathrm{bg}(r)$ then reproduces the background 
scattering length of $a_\mathrm{bg}=-450\ a_\mathrm{Bohr}$ 
\cite{Donley02}. 
In accordance with Ref.~\cite{Mies00} we model the closed channel 
potential as $V_\mathrm{cl}(B,r)=V_\mathrm{bg}(r)+E_\mathrm{cl}(B)$, where
$E_\mathrm{cl}(B)$ follows the dependence of the energy difference 
of the corresponding Zeeman hyperfine levels on the magnetic field. We use 
$h^{-1}\partial E_\mathrm{cl}/\partial B=-34.6\ $MHz/mT. In this simplified
model $V_\mathrm{cl}(B,r)$ supports only two vibrational states. We assume the 
excited one to be the resonance state $\phi_\mathrm{res}$  
which thus satisfies $[-\hbar^2\nabla^2/m+V_\mathrm{cl}]\phi_\mathrm{res}=
E_\mathrm{res}\phi_\mathrm{res}$. We have chosen the off diagonal potential
as \cite{Mies00} $W(r)=\beta\exp(-r/\alpha)$ with 
$\beta/k_\mathrm{B}=38.5\ $mK and $\alpha=5\ a_\mathrm{Bohr}$, which gives a 
width of the resonance of $\Delta B=1.1\ \mathrm{mT}$ \cite{Donley02}. 
With this choice of parameters the model also produces the correct
shift between the resonance position $B_0$ and the magnetic 
field strength $B_\mathrm{res}$, where the resonance state crosses
the dissociation threshold of $V_\mathrm{bg}(r)$
($E_\mathrm{res}(B_\mathrm{res})=0$). For the $^{85}$Rb resonance
at $B_0=15.49\ $mT this shift is negative and of the order of
$B_0-B_\mathrm{res}=-0.9 \ $mT,
so that the multi-channel bound state $\phi_\mathrm{b}$ persists  
when the resonance state has crossed the threshold. The corresponding wave 
functions and negative binding energies are obtained from Eq.~(\ref{H2B}) 
by $H_\mathrm{2B}\phi_\mathrm{b}=E_\mathrm{b}\phi_\mathrm{b}$ and the 
scattering length is approximated by 
\begin{align}
a(B)=a_\mathrm{bg}\left(1-\frac{\Delta B}{B-B_0}\right).
\label{scattlength}
\end{align}

\begin{figure}[tb]
\begin{center}
\includegraphics[width=0.45\textwidth]{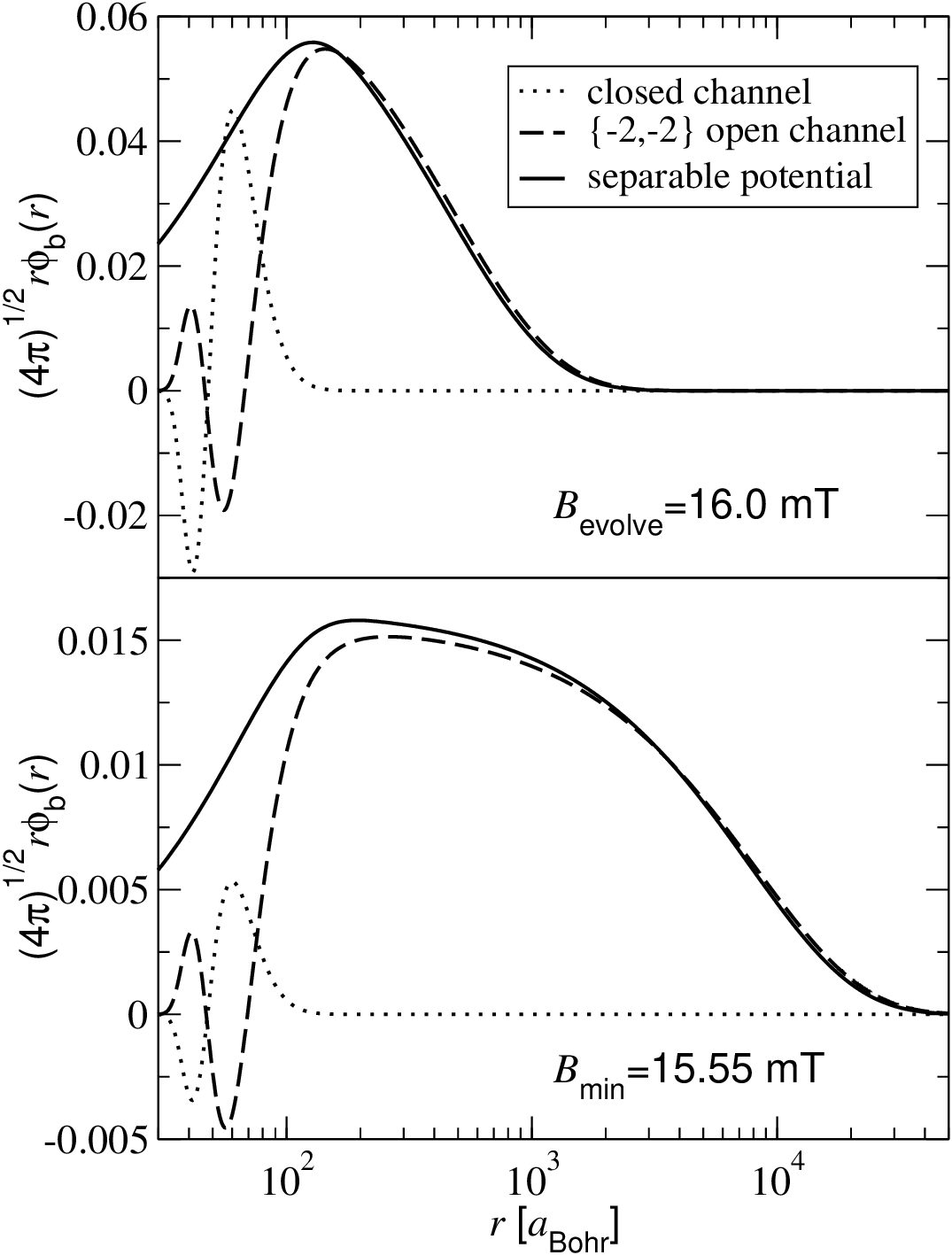}
\end{center}
\caption{Coupled-channel bound states 
at the magnetic field strengths of $B_\mathrm{evolve}=16.0\ $mT and 
$B_\mathrm{min}=15.55\ $mT.
The radial coordinate is given on a logarithmic scale.
The dotted (dashed) curves indicate the closed ($\{-2,-2\}$ open) channel 
components. 
The solid curves are the corresponding bound state wave functions of the 
separable potential in Ref.~\cite{KGB02}. The separable potential wave 
functions agree with the $\{-2,-2\}$ open channel components of the coupled
channels bound states at distances large compared to the van der Waals length 
of $\frac{1}{2}(mC_6/\hbar^2)^{1/4}=82\ a_\mathrm{Bohr}$.}
\label{fig:phib}
\end{figure}

Figure \ref{fig:phib} shows the bound state 
$\phi_\mathrm{b}$ at the magnetic field strengths of
$B_\mathrm{evolve}=16.0\ $mT and $B_\mathrm{min}=15.55\ $mT. 
While $B_\mathrm{evolve}$ corresponds to the magnetic field during the
evolution period of the pulse sequence in Fig.~\ref{fig:pulse}, 
$B_\mathrm{min}$ is the magnetic field strength closest to resonance. 
Both molecular 
states in Fig.~\ref{fig:phib} are dominated by their $\{-2,-2\}$ open channel 
component which contributes 95.3\% (99.9\%) of the total probability density 
at $B_\mathrm{evolve}$ ($B_\mathrm{min}$). For both magnetic field strengths 
the small closed channel component of $\phi_\mathrm{b}$ assumes the form of 
the  resonance state $\phi_\mathrm{res}(r)$. Close to the resonance the bond 
length of $\phi_\mathrm{b}$ (mean internuclear distance) 
is on the order of
$\langle r\rangle=a/2$. One obtains $\langle r\rangle=326\ a_\mathrm{Bohr}$ 
($a=521\ a_\mathrm{Bohr}$ from Eq.~(\ref{scattlength})) at  
$B_\mathrm{evolve}$ and $\langle r\rangle=4255\ a_\mathrm{Bohr}$
($a=7800\ a_\mathrm{Bohr}$ from Eq.~(\ref{scattlength})) at $B_\mathrm{min}$. 
At the lowest magnetic field strengths in the pulse sequence in 
Fig.~\ref{fig:pulse} the spatial extent of the molecules thus becomes 
comparable with the mean distance between the condensate atoms in the JILA 
experiments \cite{Donley02}, which is of the order of 
$n_0^{-1/3}=0.64\ \mu\mathrm{m}=12000\ a_\mathrm{Bohr}$. 
Under these conditions, the dilute gas parameter $\sqrt{n_0a^3}$ is 
comparable to 1, and one can no longer identify a particular pair of atoms in 
the gas as a molecule in the state $\phi_\mathrm{b}$ because its molecular 
wave function would overlap with other gas atoms. One and the same atom could 
thus contribute to several diatomic molecules. After the pulse sequence and 
during the evolution period, however, the gas is weakly interacting 
($n_0a^3\ll 1$) and diatomic molecular bound states are sufficiently confined 
in space for the number of molecules to be a meaningful quantity. 

We shall illustrate the resulting remarkable evolution of the molecular 
fraction in the JILA experiments \cite{Donley02} with the quantum 
mechanical many-body description developed in 
Ref.~\cite{KGB02}
for a homogeneous gas with the experimental mean density of 
$n_0=3.9\times 10^{12}\ \mathrm{cm}^{-3}$. The underlying cumulant approach 
\cite{Fricke96,Koehler02}
is based on the microscopic many-body Hamiltonian \cite{KGB02}
and systematically decouples the exact hierarchy of dynamic equations for 
quantum correlation functions at any desired degree of accuracy. 
According to Ref.~\cite{KGB02} the dynamics of the condensate mean 
field $\Psi$, the molecular fraction of the gas, and the burst of correlated 
pairs of comparatively hot atoms 
\cite{Donley02,KGB02} 
are all determined by a single nonlinear Schr\"odinger equation. In the 
homogeneous gas limit this has the form: 
\begin{align}
i\hbar\frac{\partial}{\partial t}\Psi(t)=
-\Psi^*(t)\int_{t_0}^\infty d\tau \
\Psi^2(\tau)\frac{\partial}{\partial \tau}
h(t,\tau).
\label{NLS}
\end{align}
Here $t_0$ is the initial time of the pulse (see Fig.~\ref{fig:pulse}) 
and $h(t,\tau)$ is the coupling 
function that accounts for the interactions of the gas atoms. In 
Ref.~\cite{KGB02} we have described the binary collision dynamics in terms 
of a single separable potential $V(t)$ that mimics the low energy scattering 
properties of the two component coupled-channel Hamiltonian (\ref{H2B}) and, 
at each magnetic field strength $B(t)$, the long range of the $\{-2,-2\}$
component of the coupled-channel bound state (see Fig.~\ref{fig:phib}). 
The coupling function includes the entire two-body 
time development operator 
$U_\mathrm{2B}(t,\tau)$ 
\cite{KGB02}
and is given by
\begin{align}
h(t,\tau)=(2\pi\hbar)^3\theta(t-\tau)
\langle 0|V(t)U_\mathrm{2B}(t,\tau)|0\rangle.
\label{hcoupling}
\end{align}
Here $\theta(t-\tau)$ is the step function that yields 1 at $t>\tau$ and 
vanishes elsewhere, and $|0\rangle$ is the zero momentum plane wave. In this 
treatment $U_\mathrm{2B}(t,\tau)$ accounts, in a unified manner, for both 
bound and free molecular states formed during the pulse sequence.

In Ref.~\cite{KGB02} we have determined the number of diatomic molecules in a 
weakly interacting gas in terms of the quantum mechanical observable 
\cite{Dollard73} 
that counts, for each of the $N(N-1)/2$ pairs 
of atoms in a gas with $N$ atoms, the overlap between the molecular bound 
state wave function and the many-body state. 
We have shown explicitly in Ref.~\cite{KGB02} that,
despite the $N^2$ scaling behavior of the number of pairs of atoms, our 
microscopic many-body description predicts the number of diatomic molecules 
to be smaller than $N/2$, as soon as $n_0a^3\ll 1$. 
According to Ref.~\cite{KGB02}, for a homogeneous gas, the amplitude of the 
density of the molecular fraction is given by
\begin{align}
\Psi_\mathrm{b}(t)=\frac{1}{\sqrt{2}}\int d^3r \ 
\phi_\mathrm{b}^*(r)\left[\Phi(\mathbf{r},t)+\Psi^2(t)\right]. 
\label{psib}
\end{align}
Here $t$ is the time at which 
the molecules are detected (and thereby destroyed), and
\begin{align}
\Phi(\mathbf{r},t)=-(2\pi\hbar)^\frac{3}{2}\int_{t_0}^t d\tau \
\Psi^2(\tau)
\frac{\partial}{\partial \tau}
\langle\mathbf{r}|U_\mathrm{2B}(t,\tau)|0\rangle
\label{pairfunction}
\end{align}
is the pair function \cite{KGB02,Koehler02}.
The densities of the molecular fraction and the atomic condensate 
are then obtained by $n_\mathrm{b}(t)=|\Psi_\mathrm{b}(t)|^2$ and
$n_\mathrm{c}(t)=|\Psi(t)|^2$, respectively. 

\begin{figure}[tb]
\begin{center}
\includegraphics[width=0.45\textwidth]{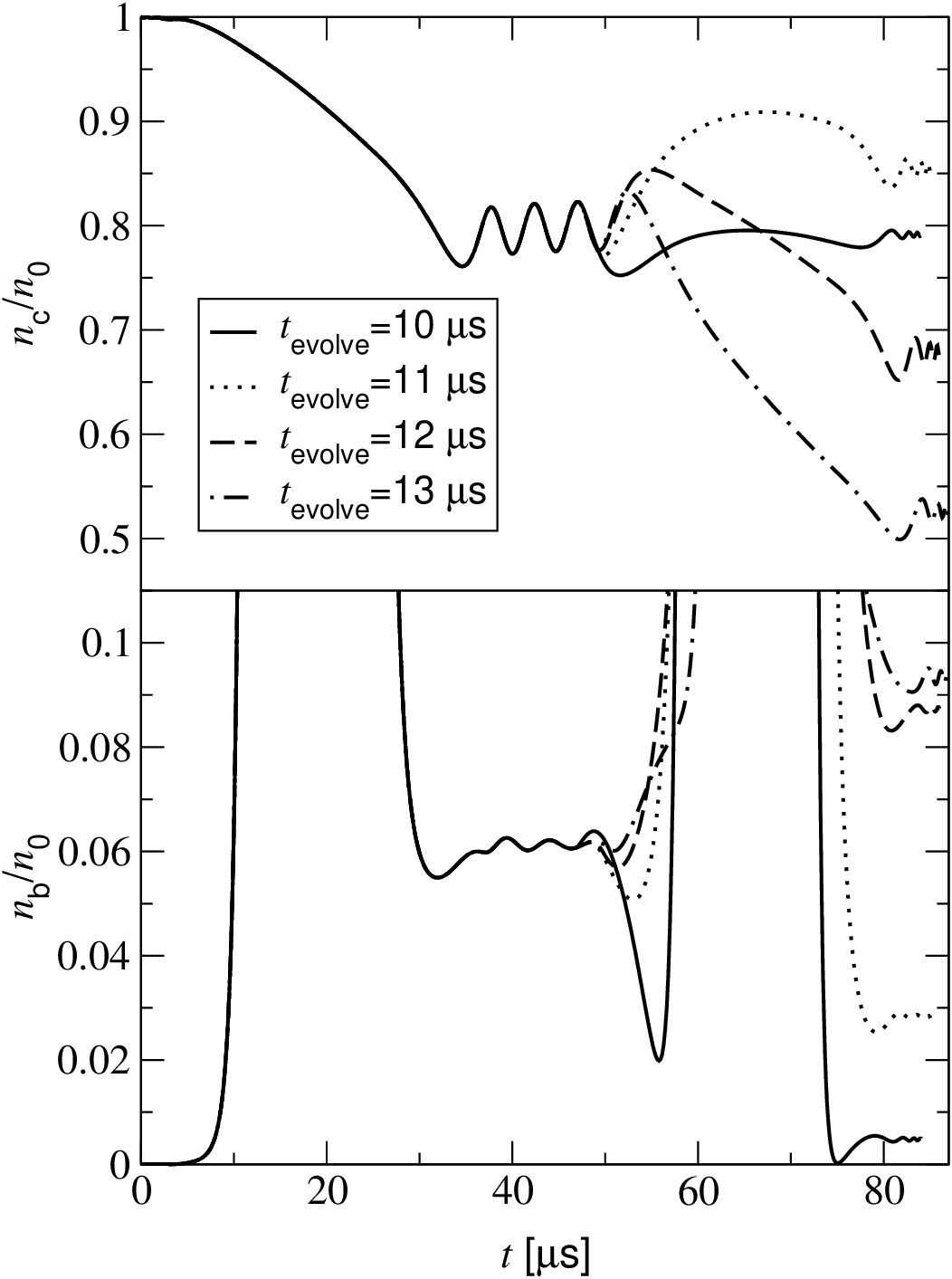}
\end{center}
\caption{Evolution of the densities of the atomic condensate $n_\mathrm{c}$ 
and the molecular fraction $n_\mathrm{b}$ for different evolution times
of the magnetic field pulse sequence in Fig.~\ref{fig:pulse}
($t_\mathrm{evolve}=10,11,12,13\ \mu\mathrm{s}$). The initially pure 
homogeneous $^{85}$Rb condensate has a density of 
$n_0=3.9\times 10^{12}\ \mathrm{cm}^{-3}$.}
\label{fig:nboft}
\end{figure}

Figure \ref{fig:nboft} shows the evolution of the atomic condensate density
and the molecular fraction in a homogeneous gas with a density of 
$n_0=3.9\times 10^{12}\ \mathrm{cm}^{-3}$ for different evolution times of the 
magnetic field pulse sequence in Fig.~\ref{fig:pulse}. As explained above,
the molecular fraction is physically meaningful only at the 
beginning ($t_0=0$) and end ($t_\mathrm{fin}=84,85,86,87\ \mu\mathrm{s}$ 
in Fig.~\ref{fig:nboft}) of the pulse sequence as well as in the evolution 
period (beginning at $37\ \mu\mathrm{s}$ in Fig.~\ref{fig:nboft}). 
At $t_0=0$ the gas is prepared as a pure atomic condensate
so that $n_\mathrm{b}(t_0)=0$. The final densities 
$n_\mathrm{c}(t_\mathrm{fin})$ and $n_\mathrm{b}(t_\mathrm{fin})$ 
exhibit the experimentally observed \cite{Donley02} pronounced oscillations 
depending on the evolution time $t_\mathrm{evolve}$ 
(see Fig.~\ref{fig:pulse}), with a frequency 
$|E_\mathrm{b}(B_\mathrm{evolve})|/h$ 
and a relative phase shift depending on the relative 
magnitudes of all components of the gas.
The amplitude of the oscillation in $n_\mathrm{b}(t_\mathrm{fin})$ is 
about twice the near constant molecular fraction  
in the evolution period of $n_\mathrm{b}/n_0=6\%$. 
The small oscillations in $n_\mathrm{c}$ and $n_\mathrm{b}$ during the 
evolution period in Fig.~\ref{fig:nboft} indicate the small 
($n_0[a(B_\mathrm{evolve})]^3=8\times 10^{-5}$)
but finite overlap between the atomic condensate and the molecular fraction.
At the lowest value of the field pulses in Fig.~\ref{fig:pulse} one obtains
$n_0[a(B_\mathrm{min})]^3=0.27$ and the observable that 
determines the number of bound pairs in a weakly interacting gas even yields 
$n_\mathrm{b}>n_0/2$ (not shown explicitly in Fig.~\ref{fig:nboft}). 
This clearly indicates the significant overlap between the molecular wave 
function of a pair of atoms with its surrounding gas atoms. The gas is then 
described by a strongly correlated non-stationary many-body state.

The entire dynamics of the different components of the gas can be understood 
in an intuitive way when the linear ramps of the two field pulses in 
Fig.~\ref{fig:pulse} are idealized as sudden switches of the magnetic field 
strength. The first field pulse then shifts the virtually uncorrelated 
initial condensate into a many-body state with the crucial binary 
correlations. The overlap of this many-body state with the multi-channel 
bound state of the pairs of atoms at the field strength 
$B_\mathrm{evolve}$ determines the molecular fraction in the evolution 
period. The overlap with the corresponding excited binary scattering states 
provides a first burst fraction \cite{Donley02,KGB02}
of correlated pairs of comparatively hot atoms. In the evolution period the 
atomic condensate and the molecular component are virtually orthogonal and 
evolve 
coherently. At the end of the evolution period the 
difference in phase between the amplitude of the atomic condensate and the 
molecular fraction is 
$\Delta\varphi=E_\mathrm{b}(B_\mathrm{evolve})t_\mathrm{evolve}/\hbar$. 
The second field pulse gets both components to overlap and, thereby, probes 
their phase difference. The second burst fraction as well as the final 
atomic condensate and molecular component thus exhibit an interference 
depending on $\Delta\varphi$. As the field pulses are chosen
as mirror images in Fig.~\ref{fig:pulse}, at constructive interference, 
$n_\mathrm{b}(t_\mathrm{fin})$ can be expected to result in about twice the 
molecular density in the evolution period. 

We have shown in this letter that highly excited diatomic bound states 
produced in recent experiments 
\cite{Donley02}
are characterized by a large spatial extent that by far exceeds the size of
all known ground state molecules 
\cite{Grisenti00}. 
The corresponding $^{85}$Rb$_2$ wave 
functions are strongly dominated by their $\{-2,-2\}$ open channel component. 
We have analyzed the non-adiabatic association mechanism of 
Ref.~\cite{Donley02} on the basis of a microscopic quantum mechanical 
many-body description of the gas. Our predicted molecular fraction of $6\%$ 
in the evolution period of the pulse sequence \cite{Donley02}
exceeds the results of Ref.~\cite{Kokkelmans02} by more than an order of 
magnitude. This provides a significantly better perspective for 
proposed experimental schemes to detect the molecules. Corrections imposed 
by the atom trap 
\cite{KGB02}
do not affect the orders of magnitude reported here. Their systematic study
in connection with possible improvements of molecular production schemes 
will be subject to future work.

We thank Eleanor Hodby, Neil Claussen, Simon Gardiner and Bill Phillips for 
inspiring discussions. This work was supported by the U.K.~EPSRC
(T.K.). K.B.~is a Royal Society Wolfson Merit Award holder.


\begin{thebibliography}{99}
\bibitem{molfield}
References \cite{Kokkelmans02,Mackie02} separate out physical observables 
associated with diatomic bound states in terms of a phenomenological 
quantum field that is included in the Hamiltonian. The molecular 
fraction reported in \cite{Kokkelmans02} is less than $0.5\%$ throughout 
the evolution period of the pulse sequence and rises steeply at the 
final ramp. The results in this letter will exceed these predictions by
more than an order of magnitude.
\bibitem{asymp}
This is the long range asymptotic form of the free Green's function
evaluated at the near resonant binding energy $E_\mathrm{b}=-\hbar^2/ma^2$
(see, e.g., \cite{KGB02}).
\end{thebibliography}
\end{document}